\documentclass[5p,number,sort&compress]{elsarticle}
\usepackage{lineno}
\usepackage{natbib}
\usepackage{hyperref}
\usepackage{graphics}
\usepackage{amsfonts}
\usepackage{amssymb,amsmath}
\usepackage{upgreek}

\begin{document}

\begin{frontmatter}

\title{Acoustic Calibration for the KM3NeT Pre-Production Module}

\author[ecap]{A.~Enzenh\"ofer\corref{cor1}}
\ead{alexander.enzenhoefer@physik.uni-erlangen.de}
\ead[url]{http://www.acoustics.physik.uni-erlangen.de}

\author{on behalf of the KM3NeT consortium}
 
\cortext[cor1]{Corresponding author}

\address[ecap]{Friedrich-Alexander-Universit\"at Erlangen-N\"urnberg, Erlangen Centre for Astroparticle Physics, Erwin-Rommel-Str.\,1, D-91058 Erlangen, Germany}

\begin{abstract}

The proposed large scale Cherenkov neutrino telescope KM3NeT will carry photo-sensors on flexible structures, the detection units.
The Mediterranean Sea, where KM3NeT will be installed, constitutes a highly dynamic environment in which the detection units are constantly in motion.
Thus it is necessary to monitor the exact sensor positions continuously to achieve the desired resolution for the neutrino telescope.
A common way to perform this monitoring is the use of acoustic positioning systems with emitters and receivers based on the piezoelectric effect.
The acoustic receivers are attached to detection units whereas the emitters are located at known positions on the sea floor.
There are complete commercial systems for this application with sufficient precision.
But these systems are limited in the use of their data and inefficient as they were designed to perform only this single task.
Several working groups in the KM3NeT consortium are cooperating to custom-design a positioning system for the specific requirements of KM3NeT.
Most of the studied solutions hold the possibility to extend the application area from positioning to additional tasks like acoustic particle detection or monitoring of the deep-sea acoustic environment.
The KM3NeT Pre-Production Module (PPM) is a test system to verify the correct operation and interoperability of the major involved hardware and software components developed for KM3NeT.
In the context of the PPM, alternative designs of acoustic sensors including small piezoelectric elements equipped with preamplifiers inside the same housing as the optical sensors will be tested.
These will be described in this article.

\end{abstract}


\end{frontmatter}

%\linenumbers

\section{Introduction}
\label{sec:introduction}

KM3NeT \cite{KM3NET} is a planned large scale neutrino telescope which will be located in the deep-sea environment of the Mediterranean.
It will consist of flexible structures, the detection units (DU), anchored to the sea bed to instrument a water volume exceeding $1\,\textrm{km}^3$.
This size is necessary to detect high energy neutrinos as the expected flux of these particles strongly decreases with their energy.
KM3NeT is designed as a deep-sea water Cherenkov neutrino telescope following the predecessor experiments ANTARES\footnote{Astronomy with a Neutrino Telescope and Abyss environmental RESearch} \cite{ANTARES}, NEMO\footnote{NEutrino Mediterranean Observatory} \cite{NEMO} and NESTOR\footnote{Neutrino Extended Submarine Telescope with Oceanographic Research} \cite{NESTOR}.
The experience gained in these projects benefit all activities related to KM3NeT.
The detection principle used in Cherenkov neutrino telescopes is based on the detection of light emitted by charged particles travelling faster than the light group velocity in its surrounding dielectric medium.
The underlying mechanism is well described by the Cherenkov effect.
This method requires sensitive optical sensors with inter sensor spacings of the order of 100\,m corresponding to the absorption length of the relevant light wavelength.
Neutrinos, being neutral particles, do not emit Cherenkov radiation.
Their detection is based on the detection of a muon generated in a neutrino interaction.
The arrival times of light signals ``seen'' by the optical sensors can be used to reconstruct the muon track from which the neutrino track can be derived.
The relative arrival time of a photon at the photo-sensor can be reconstructed with a precision of the order of 1\,ns, see e.g.~\cite{time_cal}.
It is thus mandatory for this track reconstruction to know the exact position of the optical sensor to a precision of about 20\,cm (the refractive index is about 1.33 in sea water).

The detection units are exposed to a highly dynamic deep-sea environment with sea currents varying both in velocity and direction and thus the detection units are constantly changing their positions/orientations compared to the undeflected equilibrium position.
This necessitates a continuous monitoring of the exact sensor positions.
The monitoring with the help of electromagnetic waves is impractical due to the large signal absorption and attenuation in sea water which implies a huge amount of monitoring devices to cover several $\textrm{km}^3$ volume.
Instead acoustic waves constitute a well established way to perform the position calibration of the detector.
The attenuation length of sound waves in sea water is of the order of 1\,km (at 25\,kHz, frequency dependent), signals from an emitter at the sea floor can thus be received at the top of a DU.
The viability of acoustic calibration strongly depends on the feasibility to emit well defined signals and to detect these pressure differences (sound waves) besides a high ambient pressure of about $10^7$\,Pa per 1000\,m water depth.
In addition, the calibration signals should be long ranged to allow for enhanced sensor spacings but with sufficient protection for the biosphere, e.g.~lowest possible signal amplitudes.
A set of dedicated emitters and receivers are necessary for this method.
The emitters will be located at fixed positions on the sea floor to ensure a well defined position over the whole operation period of the detector.
The anchors of the detection units are well suited positions for emitters as there will be no need for additional infrastructures to power and control these devices.
The exact position of the emitters has to be determined only once, e.g.~from a ship above the detector using an emitter with known position.
The receivers of the acoustic signals will be attached to the detection units to determine their positions relative to the different emitters.
This is done by triangulation of the different signal arrival times at the detection units with respect to the known emitter positions and emission times.
It is not necessary to equip each optical sensor with an acoustic receiver as it is possible to interpolate the sensor positions between two distant acoustic receivers using a model of the shape of the DU, cf.~the positioning system in ANTARES \cite{positioning}.
It is possible to use either complete commercial systems or to combine different emitters and receivers to a dedicated system.
Latter systems offer more flexibility concerning signal amplitudes, frequencies and signal shapes whereas commercial ones are characterised by a sophisticated and reliable design.
The use of a dedicated system is preferred as it allows for more diverse applications and studies.
The devices used for this kind of applications, both for emission and reception, are mostly based on the piezoelectric effect.
These devices are capable of withstanding the high ambient pressure of the deep sea while still offering a simple and reliable design without mechanical parts.
The long term stability of piezoelectric ceramics, hereafter called piezos, ensures the operability of the acoustic calibration system over the whole lifetime of the detector.
In this article a short overview is given of different possible types of piezos to be used for acoustic calibration as well as a more detailed description of a new development in this field at the ECAP\footnote{Erlangen Centre for Astroparticle Physics} \cite{ECAP}.
Some of these devices will be tested on the Pre-Production Module (PPM) of KM3NeT in order to find the best solution to monitor the positions of the detection units.

\section{Sensor types}
\label{sec:sensor_types}

All sensors that are considered here comprise a piezo and are thus based on the piezoelectric effect.
If such a piezoelectric crystal is exposed to external pressure/force variations, it accumulates electrical charge on its surfaces (depending on the polarisation of the crystal and the direction of the variation).
The resulting difference in potential between those surfaces can be measured as a voltage signal.
This voltage signal is proportional to the applied force.
The voltage signal resulting from the deformation is very small and has to be amplified prior to its analysis.
Typical reception sensitivities for commonly used piezos are of the order of $-200\,\text{dB re}\,1\frac{\text{V}}{\upmu \text{Pa}}$.
This effect is reversible and in this way a piezo can also be used as acoustic emitter if it is exposed to an external electrical field.
A sufficiently large high voltage signal has to be applied to the emitter to generate a useful signal amplitude in this case.
Typical emission sensitivities for commonly used piezos are of the order of $140\,\text{dB re}\,1\frac{\upmu \text{Pa}}{\text{V}}\,@\,1\,\text{m}$.
The sensitivity and directivity of the piezo strongly depends on the used material as well as on its geometry and polarisation.
There is a variety of ultrasonic devices designed especially for acoustic calibration as well as for other different applications.

\begin{figure}
\begin{center}
\includegraphics[width=\linewidth]{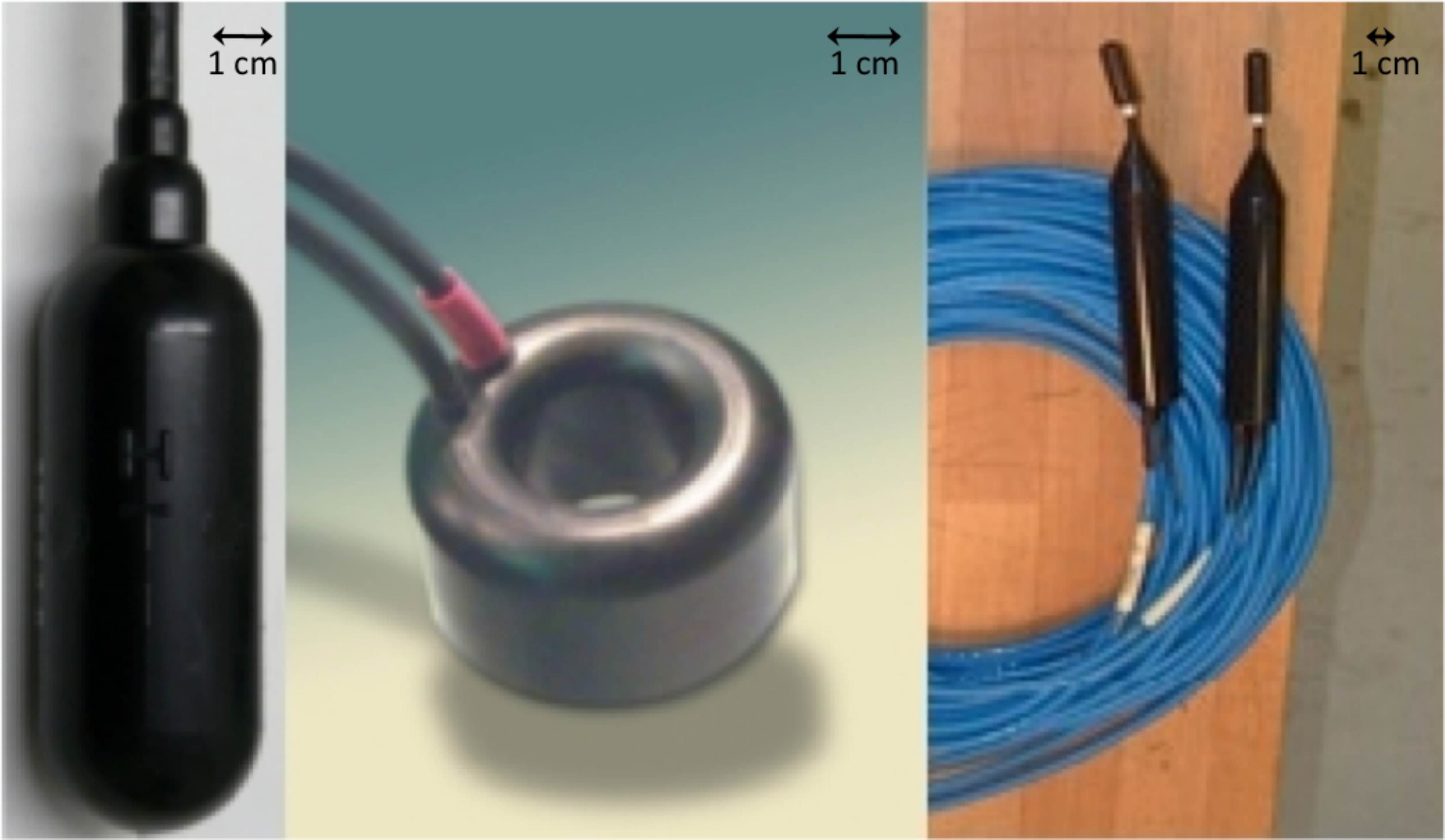}
\caption[Different sensor types]{A short description of these three sensor types is given in the text.}
\label{fig:Sensors}
\end{center}
\end{figure}

Figure \ref{fig:Sensors} shows an exemplary set of acoustic sensors.
The first one on the left side is a hydrophone from High Tech, Inc.~\cite{hti} as used in the AMADEUS\footnote{ANTARES Modules for Acoustic DEtection Under the Sea} test setup \cite{amadeus_nimA}.
These hydrophones comprise a hollow piezo cylinder and a preamplifier moulded together in polyurethane.
They are almost equally sensitive over a frequency range from 10\,kHz to 50\,kHz.
This is the frequency range of interest for acoustic neutrino detection.
These hydrophones are used for investigations of acoustic particle detection, positioning (and marine science).
The acoustic sensor in the middle, cf.~Figure \ref{fig:Sensors}, is a ``Free Flooded Ring'' (FFR) from SensorTech \cite{ffr}.
This transceiver can be used either as emitter or as receiver as it is not equipped with a preamplifier.
This sensor type is studied at IGIC-UPV\footnote{IGIC-Universitat Poilt\`ecnica de Val\`encia}.
It is planned to use this devices as emitters for the KM3NeT acoustic positioning system.
Its use as receiver is also under investigation.
In the latter case the FFRs are equipped with a preamplifier to detect small pressure variations.
The third sensor type seen in Figure \ref{fig:Sensors} is a hydrophone from SMID \cite{smid} equipped with a preamplifier.
This hydrophone was developed in cooperation with INFN\footnote{Istituto Nazionale di Fisica Nucleare} and is designed to work in the frequency range from 10\,Hz to 70\,kHz.
These sensors are used in the NEMO Phase \MakeUppercase{\romannumeral 2} framework and are also under consideration for the use in the KM3NeT detector.
The acoustic monitoring of the deep sea environment or long-run monitoring of the sea in general is a multidisciplinary task which also can be adressed with these types of sensors.
Almost all of these devices and their electronics are moulded into a single housing, e.g.~in polyurethane, to protect them from the sea water, the high ambient pressure and other environmental influences.
Polyurethane is a frequently-used material for underwater acoustics as its acoustic impedance can be matched with the one of the surrounding water.
This ensures the best performance for emission and reception while still protecting the device.
Another possibility to protect the acoustic receiver is presented in the next section.

\section{Opto-Acoustical Modules (OAMs)}

\begin{figure}
\begin{center}
\includegraphics[width=\linewidth]{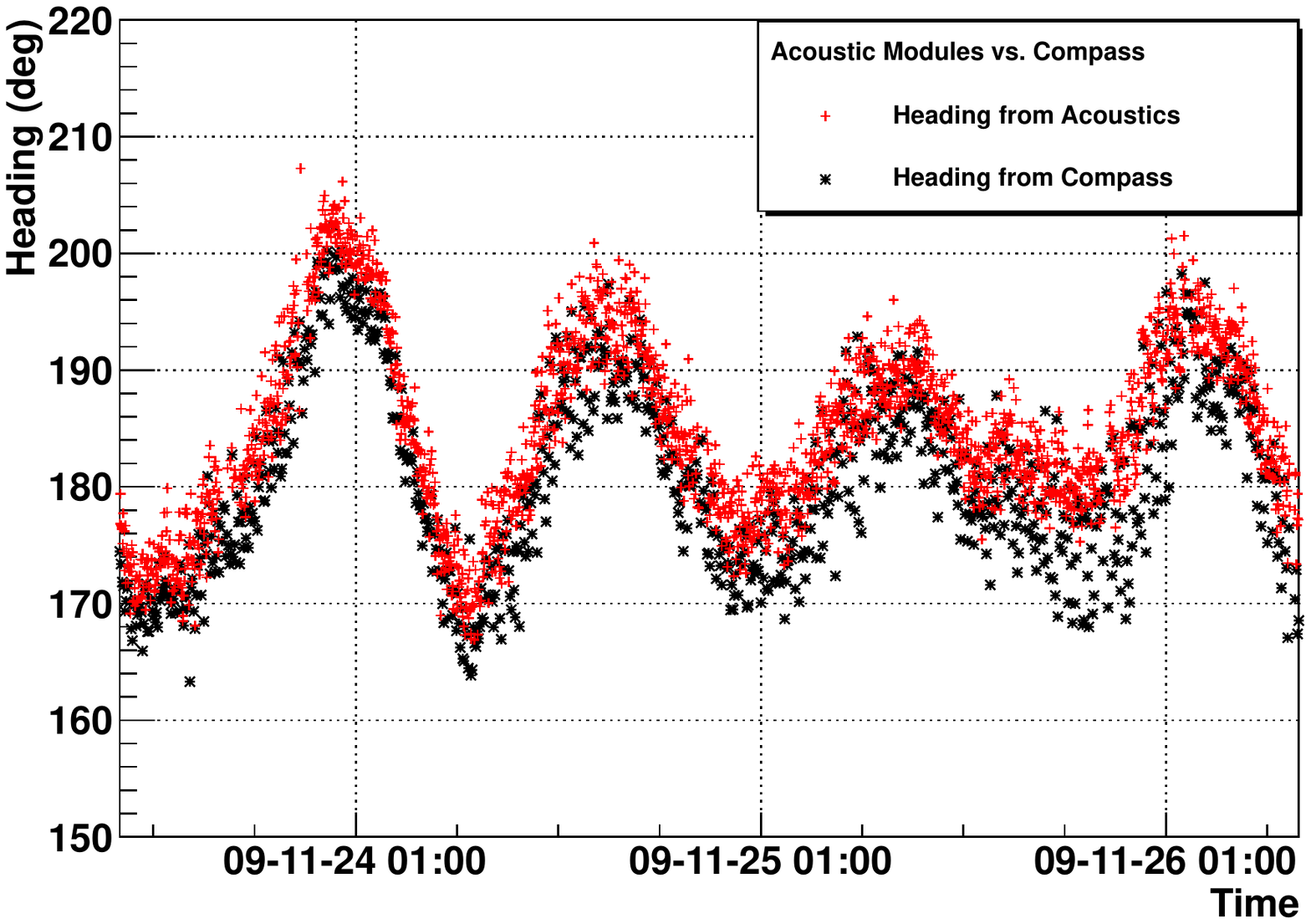}
\caption[Comparison of heading measurements]{The graph depicts heading measurements carried out for an ANTARES structure holding AMs. The data obtained from a dedicated compass board are in good agreement to the heading reconstructed with the AM data.}
\label{fig:heading}
\end{center}
\end{figure}

The so-called Acoustic Modules (AMs) deployed in the AMADEUS test setup represent a different approach to the aforementioned moulding of acoustic sensors.
These modules consist of two bare piezos glued to the inside of pressure resistant glass spheres, identical to those used for optical modules.
Each piezo is connected to a preamplifier which is located inside a copper tube together with the piezo to shield the resulting sensor from electromagnetic influences.
The major difference compared to the use of dedicated receivers is the possible combination of acoustic sensor and optical sensor inside the same housing.
The resulting Opto-Acoustical Module (OAM) would combine all features of a standard optical module, the key building blocks of the detection units, with the ability to determine its position with respect to a set of dedicated emitters without the need for additional calibration devices.
This reduces the required underwater connectors and feedthroughs to a minimum as one connection is sufficient to connect the OAM.
Underwater connectors and feedthroughs constitute potential points of failure with water ingress in the worst case.
There is no need to further protect the acoustic sensor against environmental influences as it is perfectly protected by the glass sphere.
The results obtained with the AMs prove the principal concept to operate this type of acoustic sensor.
A result comparing exemplary the reconstructed heading using AMs with the one obtained through dedicated compass devices is shown in Figure \ref{fig:heading}.
The relative difference between both methods is of the order of $1^{\circ} - 2^{\circ}$ which is of the same order of the accuracy given by the compass device manufacturer ($\sim 1^{\circ}$).

\begin{figure}
\begin{center}
\includegraphics[width=\linewidth]{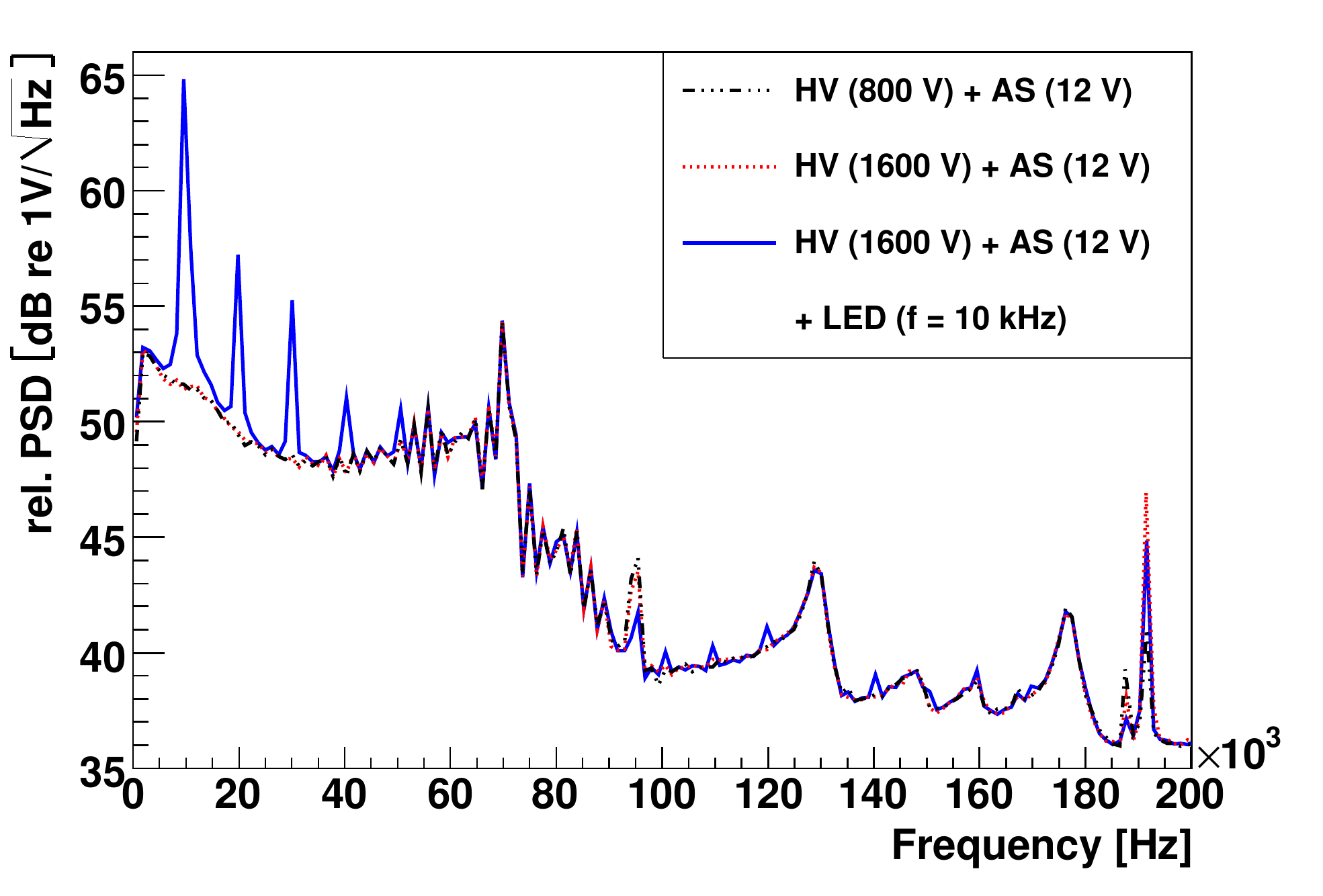}
\caption[First prototype test]{The figure shows the results of noise measurements for a first OAM prototype and different settings given in the legend. This test in a simplified setup shows some prominent features. These can be attributed to the poor shielding of the acoustic sensor as well as to the power supply. ``HV'' shows the high voltage between cathode and anode of the PMT. ``AS'' stands for the acoustic sensor which was powered by 12\,V and ``LED'' gives the frequency of a flashing LED inside the test setup.}
\label{fig:old_design}
\end{center}
\end{figure}

The results prove the possibility to determine the position of the AMs to a sufficient precision.
A study of the angle dependence of a piezo glued to a glass sphere can be found in \cite{diplom_ae}.
The biggest obstacle to a combined sensor module is expected from the high voltage necessary to operate a photomultiplier inside the module.
The high voltage itself or its transformation from a lower supply voltage might lead to a ``noisy'' environment interfering with the signal path of the acoustic device.
On the other hand the acoustic device with its amplifiers might influence the PMT operation.
The latter case is unlikely and was not observed so far due to the low voltage ($\sim$ 3.3\,V to 5.0\,V) necessary to power the acoustic sensor but any parasitic influence has to be eliminated.

\begin{figure}
\begin{center}
\includegraphics[width=4cm]{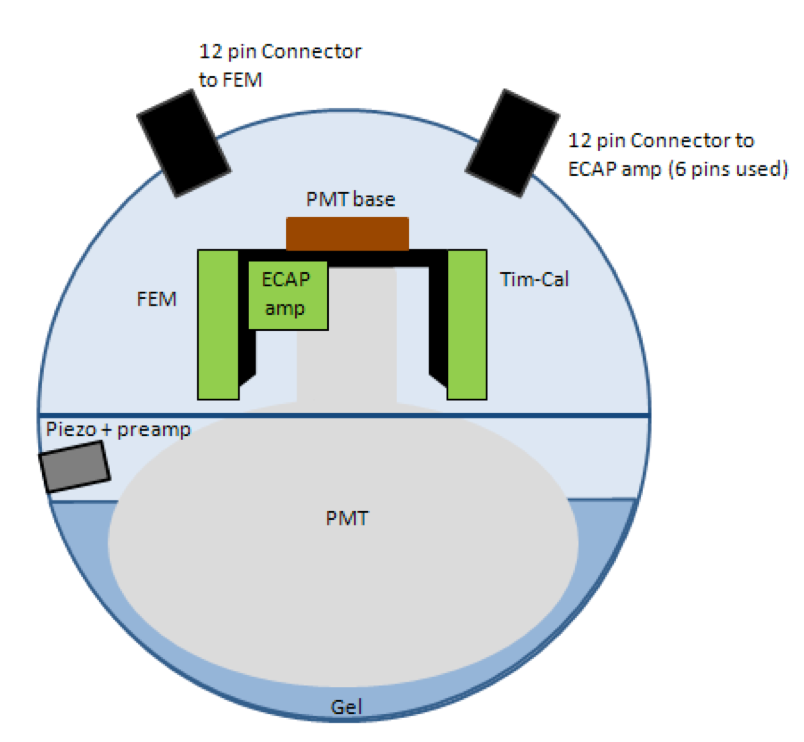}
\includegraphics[width=3.5cm]{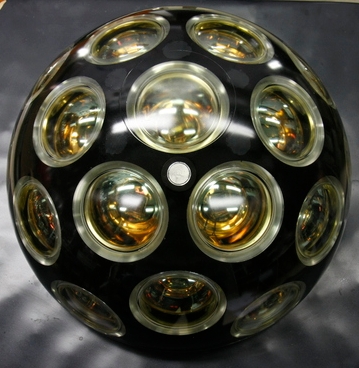}
\caption[Design options for a detection unit]{The figure on the left shows a schematic of a single PMT optical module equipped with an acoustic sensor to be used in NEMO Phase \MakeUppercase{\romannumeral 2}. The picture on the right shows the exterior of a prototype multi PMT detection unit. The PMTs are clearly visible as well as a small circle, the metallised piezo surface.}
\label{fig:du_options}
\end{center}
\end{figure}

First tests, with results shown in Figure \ref{fig:old_design}, demonstrated that PMTs and piezos can be operated within a single sphere.
Some interference of the PMT operation with the piezo was observed.
This can be attributed to the very poor electromagnetic shielding of the first prototypes (no electromagnetic shielding was applied) and the pulsed LED (connected to the same power supply as both sensors).
Some tests in cooperation with LNS-INFN\footnote{Laboratori Nazionali del Sud - Istituto Nazionale di Fisica Nucleare} \cite{lns_infn} were carried out to test the combined operation under more realistic conditions.
In this context the NEMO Phase \MakeUppercase{\romannumeral 2} infrastructure was used together with a new prototype of a piezo-preamp unit with better shielding and a different amplifier layout adapted to the operation conditions.
In this configuration no drawbacks were visible during laboratory tests with the piezo mounted in a 13$''$ sphere next to a PMT powered almost at its nominal voltage with a dark count rate of several kHz.
The small size of the acoustic sensor enables its integration into a variety of different optical module designs.
See Figure \ref{fig:du_options} for the two designs currently pursued.
The first one is the single PMT option used in the NEMO Phase \MakeUppercase{\romannumeral 2} framework and the second one is a multi PMT (31 PMTs per module) option developed at Nikhef\footnote{Nationaal instituut voor subatomaire fysica} \cite{nikhef} in Amsterdam and will be used for the PPM.

\section{Conclusions}
\label{sec:conclusions}

The acoustic calibration is a way to determine the position and the orientation of detecton units.
The choice of the appropriate acoustic devices for this task strongly depends on the intended use.
Position calibration with the required precision can be realised with each of the presented devices.
The KM3NeT Pre-Production Module provides the necessary infrastructure to test different types of sensing devices to validate the positioning results and to figure out potential incompatibilities in the final system.
The various components are developed at different KM3NeT member institutes all over Europe.
Joint tests in the different laboratories yielded a perfect compatibility of the different sensor components.
The final confirmation requires the fully-fledged PPM evaluated under real conditions.
The multidisciplinary research adressed with general purpose devices becomes more and more important as the complexity and efforts to maintain large scale detectors increase and have to be partitioned among several institutions and between different scientific subjects.
KM3NeT will serve as an observatory where many natural sciences collaborate to better understand our planet with its deep sea environment as well as our universe as a whole.

\section*{Acknowledgements}
\label{sec:acknowledgements}

The authors acknowledge the support through the EU, FP6 Contract no. 011937 and FP7 grant agreement no. 212252.

% Bibliography

\end{document}